\documentstyle[11pt,aaspp4]{article}

\def\kms{km s$^{-1}$}

\begin{document}

\title{Imaging and spectroscopy of arcs around the most luminous 
X-ray cluster RX~J1347.5-1145\altaffilmark{1}}
\altaffiltext{1}{Based on observations with the NASA/ESA \it Hubble Space 
Telescope,
\rm obtained at the Space Telescope Science Institute, which is operated
by AURA Inc under NASA contract NAS5-26555, and the  3.6-meter telescope
of the European Southern Observatory at La Silla, Chile.} 
\author{Kailash C.~Sahu\altaffilmark{2}, Richard A. Shaw\altaffilmark{2}, 
Mary Elizabeth Kaiser\altaffilmark{3} \altaffilmark{6},
Stefi A. Baum\altaffilmark{2}, Henry C. Ferguson\altaffilmark{2}, 
Jeffrey J. E. Hayes\altaffilmark{2}, 
Theodore R. Gull\altaffilmark{3}, 
Robert J. Hill\altaffilmark{4}, John B. Hutchings\altaffilmark{5}, 
Randy A. Kimble\altaffilmark{3}, Philip Plait\altaffilmark{7}, 
Bruce E. Woodgate\altaffilmark{3}}
\altaffiltext{2}{Space Telescope Science Institute, 3700 San Martin Drive, 
Baltimore, MD 21218}
\altaffiltext{3}{NASA/Goddard Space Flight Center, Code 681, Greenbelt, MD 
20771} 
\altaffiltext{4}{Hughes/STX Corporation NASA/Goddard Space Flight Center, Code 
681, Greenbelt, MD 20771}
\altaffiltext{5}{Dominion Astrophyscial Observatory, National Research Council 
of 
Canada, 5071 Saanich Rd., Victoria, B.C. V8X 4M6, Canada}
\altaffiltext{6}{Dept of Physics and Astronomy, Johns Hopkins University, 
Baltimore, 
MD 21218}
\altaffiltext{7}{Advanced Computer Concepts, NASA/GSFC, Code 681, Greenbelt, 
MD 20771}

\begin{abstract}

The cluster RX J1347.5-1145,  
the most luminous cluster in the X-ray wavelengths, was imaged with the 
newly installed Space Telescope Imaging Spectrograph (STIS)
on-board HST. Its relatively high redshift (0.451) and luminosity indicate 
that this  is one of the most massive of all known clusters. 
The STIS images unambiguously show several arcs in the cluster.
The largest two arcs ($>$5$\arcsec$ length) are symmetrically situated on 
opposite sides of the cluster, at a distance of 
$\sim$ 35 arcsec from the central galaxy. 
The STIS images also show approximately 100 faint galaxies
within the radius of the arcs whose  combined 
luminosity is $\sim 4 \times 10^{11} L_\odot$.  We also present
ground-based spectroscopic observations of the northern arc which show 
one clear emission line at $\sim$6730 \AA, which is
consistent with an identification as 
[OII] 3727 \AA, implying a redshift of 0.81 for this 
arc. The southern arc shows a faint continuum but no emission features.
The surface mass within the radius of the arcs (240 kpc), 
as derived from the gravitational 
lensing, is $\sim$6.3 $\times $ 10$^{14}$ M$_\odot$. The resultant 
mass-to-light ratio of $\sim$1200 
is higher than what is seen in many clusters but smaller than the
value recently derived for some `dark' X-ray clusters (Hattori et
al. 1997).  The total surface mass derived from the X-ray flux within 
the radius of the arcs is  $\sim$2.1 - 6.8 $\times$ 10$^{14}$ M$_\odot$, which 
implies that  the
ratio of the gravitational to the X-ray mass is $\sim$1 to 3. 
The surface {\it gas} mass within this radius is 
$\sim$3.5 $\times $ 10$^{13}$ M$_\odot$, which implies that 
at least 6\% of the total mass within this region is baryonic.

\end{abstract}

\section{Introduction}

Clusters of galaxies are the most massive gravitationally
bound systems in the Universe. They serve as important probes of the
large-scale structures and their evolution (see e.g. Bahcall, 1988).
Accurate mass determination of such clusters can place very useful constraints 
on $\Omega$ and the nature of the dark matter (White and Fabian, 1995).

RXJ 1347.5-1145 (RA(2000) = 13$^h$ 47$^m$ 30.$^s$5 Dec(2000) = -11$^\circ$ 
45$\arcmin$ 09$\arcsec$ ) is the most luminous X-ray cluster among all the
clusters observed in the ROSAT all sky survey (Schindler et al. 1995). 
Its high luminosity coupled with a redshift of 0.451 implies that this is 
one of the  most massive of all clusters known to date. 

The cluster has two optically bright galaxies at the center. 
Optical spectra for these brightest cluster members indicate
that the cluster is at a redshift of z=0.451 (Schindler et al.,
1995). Optical images 
obtained by Schindler et al. (1995) and Fischer and Tyson (1997)
show two possible bright arcs symmetrically situated at $\sim$35
arcsec away from the brightest cluster member and 
some  hint of other arcs, requiring higher spatial resolution images for
confirmation.  Spectral observations for the arcs were not available.
This paper presents high spatial-resolution images obtained with
STIS, and ground-based spectroscopic observations for the arc, which are
used to estimate the mass of the cluster.

\section{Mass-distribution from X-ray observations }

Many of the X-ray bright clusters show the presence of gravitationally
lensed arcs. From an optical follow-up study of 41 X-ray bright
clusters, Gioia and Luppino (1994) showed that compared to an
optically selected sample, the X-ray selected sample has 3 times
larger probability of showing gravitationally lensed arcs.  This
clearly indicates that the luminous X-ray clusters are massive,
perhaps the most massive of all observed objects in the universe.
Their total mass determination is clearly important, and their mass
profile may give important clues to the nature of the dark matter.

These X-ray clusters provide an opportunity not only to study these
most massive members of the universe, but also to study the properties
of background galaxies imaged as arcs.  The gravitational lensing
facilitates this study in two important ways. First, it provides a
direct way to derive the mass of the lensing cluster.  Secondly, it
increases the brightness of the background galaxy by the formation of
the arcs, thereby enabling us to study the background galaxy which
would have otherwise been too faint to be studied.

In the ROSAT band of 0.1 to 2.4 kev, the X-ray emission from RXJ1347.5-1145
extends to a radius of $\sim$4.2 arcmin, with an observed
luminosity of 7.3 $\times 10^{45}$ erg s$^{-1}$ within this radius.
The ASCA observations (2 - 10 kev) show emission out to a radius
of $\sim$6.4 arcmin, with an observed luminosity of
4.6 $\times 10^{45}$ erg s$^{-1}$ in this band. 
From the X-ray observations the mass distributions within radii of 240~kpc,
1~Mpc, and 3~Mpc have been derived.  The total surface mass within the radius
of the large arcs (240 kpc) is 2.1 $\times 10^{14} M_\odot$
(Schindler et al. 1997) to 6.8 $\times 10^{14} M_\odot$ (Allen, 1997).  
Within a radius of 1 Mpc, the gas mass is 2.0 $\times 10^{14} M_\odot$ and the
total mass, assuming hydrostatic equilibrium, is 5.8 $\times 10^{14}
M_\odot$. On the 3 Mpc scale, the gas mass is 8.9 $\times 10^{14} M_\odot$
and the total mass is 1.7 $\times 10^{15} M_\odot$.

\section{Observations}

The newly installed Space Telescope Imaging Spectrograph (STIS) on
board HST was used to get the images of RX J1347.5-1145. Observations
were taken on 25 May 1997, with the CCD detector using the clear and
the long-pass filtered apertures (Baum et al. 1996). A mosaic of 4
(50$\arcsec \times 50\arcsec$) unfiltered CCD images was taken to
encompass the field of arcs which span $>$70$\arcsec$.
The field of view of the long pass filter (28$\arcsec \times 50
\arcsec$) is smaller than the clear aperture.  As a consequence, only
the brightest central galaxy and arc 2 were imaged with this
filter. The integration times for each telescope pointing ranged from
250 to 300 seconds.

Spectroscopic observations were taken using the versatile,
high-throughput EFOSC (ESO Faint Object Spectrographic Camera) at the
ESO 3.6m telescope at La Silla, Chile. The camera has both imaging and
spectroscopic capabilities. Since the spectrograph can hold 5 gratings
simultaneously, multiple spectral resolutions and bandpasses were 
easily configured.
A log of the spectroscopic observations is provided in Table 1.

First, a direct image of the field was obtained with an exposure of 30~sec. 
The seeing at this time was
$\sim$1.5$\arcsec$. A field of view of 5.2$\arcmin \times$
5.2$\arcmin$ was obtained using the 512 $\times$ 512 pixel Tektronix CCD.
A slit of 1.5$\arcsec$ width was used for all the observations.  The
first spectra were obtained with the R300 grating, with the
slit oriented so that the
northern arc was along the length of the slit.
A single, bright  emission feature at $\sim$6730~\AA~with a faint continuum on
either side can be seen in each of the two exposures taken in this
configuration. Spectra obtained in the blue with the B300 grating
did not show any additional features.  The two galaxies 
were placed in the slit and 3 exposures, each of 20 minutes duration, 
were taken with R300 grating. Fig. 1 shows the combined spectrum for the 
northern arc and the two central galaxies.

Similar observations were taken for the southern arc.  Exposures of
the southern arc in the same grating setup (R300) however showed no
trace of emission; only a faint continuum was visible.  Spectra were
then taken with the B300 grating spanning the wavelength region 
from 3600 \AA~to 6800~\AA.
No features, except a faint continuum in the blue, are visible
despite the fact that the detection limit for the southern arc
is better than for the northern arc.  If the arc has [OII] emission at
3727 \AA, then the observations indicate that it is redshifted beyond
$\sim\,$7500 \AA \ and into a region where numerous atmospheric absorption
lines make detection difficult.
This would indicate that the redshift is higher than ~1.  However,
this object could also be a galaxy without any strong emission
features in which case the continuum of the galaxy is consistent 
with a redshift on the order of 0.8 to 1.5.

\section{Data reduction and results} 

The STIS images were analyzed using both the
STScI CALSTIS and the IDT CALSTIS calibration pipelines 
(both of which essentially employ the same procedure as mentioned above), 
and consistent results were obtained. The overscan regions were first 
subtracted from the raw CCD images. Two readouts
per HST telescope pointing position were employed to facilitate cosmic
ray rejection.  The cosmic rays were removed using an  algorithm 
which compares the two
images and iteratively replaces pixels which deviate by more than 8
$\sigma$ with an appropriately weighted minimum value of the
pixel.  Next a bias image and a dark frame are subtracted from the
data. This image is then flat fielded using a ground based calibration
flat.  The resulting image is then compared to the pipeline resident
hot pixel table ($>$0.1 count sec$^{-1}$ pixel$^{-1}$).  Any residual hot pixels
are then eliminated by visual inspection using a nearest neighbor
interpolation algorithm. 

A mosaic image was created by aligning the brightest feature in each of the
overlap regions.  The images were combined using the IRAF task
`imcombine', with a
weighted average proportional to the exposure time for the overlapping
regions, and counts normalized to a mean exposure time in the entire
mosaic image.  Fig. 2 (Plate 1) shows the resultant mosaic
image. The image shows several arcs which are marked in the figure.
The two large arcs (1 and 4), situated on opposite sides of the
central galaxy are separated by about 70 arcsec; this is one of the
largest separations seen in any cluster.  Positional and dimensional
information regarding the various arcs is given in Table 2.

The two large arcs were already seen from ground-based observations.
However, due to the lower spatial resolution of the ground-based
observations, the possibility that the arcs 
are foreground galaxies could not be 
ruled out  (Fischer and Tyson, 1997). STIS observations
unambiguously prove that these are arcs with magnifications
(which, to a first order, is the length to width ratio) ranging
from 7 to 15. There are 3 additional features
whose orientation and distance from the center are consistent with
being arcs, although the possibility that these smaller arcs may be
foreground galaxies cannot be eliminated.

The two central galaxies are clearly resolved.  
Fig. 3 (Plate 2) shows an enlarged view of the central galaxy, which shows
a clear jet-like structure on one side, and
a similar but much fainter structure on the opposite side.
These structures, which may be jets or tidal tails, are about 
0.6$\arcsec$ long, which corresponds to a length of $\sim$ 3 kpc.
The FHHM of the central and the fainter
galaxies, as derived from the average Moffait and Gaussian fit to
their structure in different position angles, is 0.6 and 0.4 arcsec,
respectively. The FWHM of the point sources in the field is
$\sim$0.085 arcsec. Derived magnitudes for the two galaxies and the arcs
are given in Table 3, which are consistent with  
those of Fischer and Tyson (1997).

The image processing software packages MIDAS and IRAF were used to
reduce the spectral data obtained from the ground.  After bias
subtraction, the frames were flat-field corrected with an average
of 5 flat-field images taken with the dome illuminated with a tungsten lamp.
The sky was taken from both sides of the spectrum for a good sky 
subtraction. The resulting one dimensional spectrum was then wavelength
and flux calibrated. A He-Ar spectral lamp was used for wavelength calibration.
For flux calibration the standard stars HD 8879, a fast rotating Be star,
and LTT 9239 were used.  In the wavelength region redward of 7500
\AA~there are several atmospheric emission lines, making it
difficult to differentiate between these emission features and the
source in this region. The resulting final spectra for the northern 
arc and the two galaxies are shown in Fig. 1.

\section{Redshifts}

\subsection{The central galaxies}

The spectra for the two central galaxies (Fig. 1) are consistent with the
lower resolution (15 \AA) spectra 
published by Schindler et al. (1995).
Compared to the eastern galaxy, the central galaxy
is brighter by $\sim$0.4 magnitudes and is also bluer. The central
galaxy shows emission features of H$\beta$ $\lambda 4861$\AA, and
[OIII] $\lambda 4959$ and $5007$\AA, and the eastern galaxy shows a
few absorption features, all of which indicate the lensing cluster to
be at z$\sim$0.451. Our higher resolution spectra also enable us to
derive a velocity dispersion of 620~\kms\ for the central galaxy.
Using the radius of the galaxy and the velocity dispersion, we derive
(Binney and Tremaine, 1987) a mass of 2.4$\times$10$^{11}$ M$_\odot$
for the galaxy.  However, this mass determination is valid only if
the radius of the line emitting region is the same as the observed
radius of the galaxy, and if the system is virialized.  If the line
emitting region is confined to a smaller region, this mass
determination is not applicable.
The mass of the line emitting region itself would be lower, but
depending on the contribution of mass outside the line emitting
region, the total mass can be higher or lower.

\subsection{The arcs}

Since the redshift of the lensing cluster is 0.45, the redshift of the
arcs must be higher. This implies that the rest wavelength of the
northern arc's emission line seen at 6730 \AA \ is less than 4638
\AA.  This eliminates the possibility that the emission line could be
either of the H$\beta$ $\lambda 4861$\AA, [OIII] $\lambda 4959$ and
5007\AA \, or the H$\alpha$ $\lambda$ 6563 \AA \ lines.  The only
strong emission lines which may be responsible are the [OII] $\lambda$
3727 \AA, CIV $\lambda$ 1550 \AA \ and Ly H$\alpha$ $\lambda$ 1216 \AA
\ lines, with corresponding redshifts
of 0.8, 3.34 and 4.53, respectively.  Although we cannot rule out
the possibility that the emission line could be either Ly$\alpha$ or
CIV, the color of the arc is inconsistent with this interpretation.
The Lyman limit in that case would fall in the middle of the B-band,
and the arc would be much redder than the B$_J$--R $ \sim 1.1$ value
observed by Fischer and Tyson (1997).

The emission line is fully consistent with the line
being the [OII] line. In that case, however, 
the H$\beta$ and [OIII] line emissions should be within
the spectral region observed from the ground. Unfortunately, all of 
these lines would be hidden beneath the 
strong atmospheric bands which abound in this region
of the spectrum. Spectra obtained with HST and STIS can provide an
unambiguous determination of the redshift of the arc by detection of
other emission lines. In our subsequent analysis, we will assume that the
redshift of the arc  
is 0.8, noting that the mass estimate is not affected by more than
30\% even if the redshift is as high as 4.

The southern arc is slightly fainter but its distance is only about 2\%
 larger than the distance to the northern arc.  Furthermore,
its center of curvature is almost centered at the bright galaxy. A
simple physical model, consistent with these data and an Einstein
radius equal to the distance from the center to the arc in the lens
plane, predicts that the redshift of this arc is between 0.7 and 1.
The lack of an emission line detection
at this wavelength suggests that either the source is a galaxy devoid
of emission lines, or the emission line is hidden beneath the
abundance of atmospheric lines.

\section{Gravitational mass and the mass-to-light ratio}

Fischer and Tyson (1997) have obtained a deep ground-based
image of the cluster with a field size of dimension (14$\arcmin$)$^2$,
which they have used for a mass-modeling of the cluster
(also see Tyson and Fischer, 1995). They detect a shear-signal in the background
galaxies in the radial range of 35$\arcsec$ to 400$\arcsec$ from the cluster 
center and estimate the redshift of the arcs to be 1.4$^{+1.4}\hskip -0.6cm
_{-0.35}$ and 1.6$^{+2.0}\hskip -0.6cm
_{-0.5}$ respectively. Here we confine ourselves to the
strong lensing regime (r $<$ 35$\arcsec$).
The fact that the two arcs on the opposite sides (nos. 1 and 4) have
 rather large magnifications
and the center of the radius of curvature of the arcs
lies close to the central galaxy makes the mass modeling rather 
straight forward. In such a case, the angular radius $\theta_E$
corresponding to  the Einstein radius of the 
lensing galaxy $R_E$, can be expressed as (Schneider, 
Ehlers and Falco, 1992; Blandford and Narayan, 1992)
$$ \theta_E^2 = {{4GMD}\over {c^2}}, D = {D_{ds} \over D_d D_s}  \eqno (1)$$
\noindent where M is the  the mass of the lensing object,
D$_d$ is the the distance to the lensing object, 
D$_{ds}$ is the the
distance from the lens to the source, and
D$_s$ is the distance from the observer to the source
(all distances being angular distances).
We can assume $\theta_E$  to be  the same as the distance from the
lensing galaxy to the arc (34.9$\arcsec$ for the northern arc). Since the
redshifts of the lensing galaxy and the arc are now known, we can
derive the projected  mass within this radius.  The resulting mass
is 6.3 x 10$^{14}$ M$_\odot$ (assuming $\Omega$ = 1, $\Lambda$ = 0
and H$_0$ = 50 \kms Mpc$^{-1}$). To derive the total luminosity, 
we measured the fluxes of all
($\sim$100) galaxies within the radius of the assumed
Einstein ring of the cluster. To derive the 
magnitudes in the CCD-clear aperture, we use the following
formula as determined from the cycle 7 calibration programs:
$$ V(Johnson) = -2.5 log(c/s) + 25.69 + 0.479(V-I) \eqno (2)$$
\noindent where c refers to counts obtained with gain=1.
Using the magnitude and color information for the two central 
CD galaxies (given in Table 3) and other galaxies in the field,
we estimate that the combined flux 
from these two galaxies is approximately 38\% of the total flux from all 
the galaxies (assuming the contribution from even fainter galaxies is small). 
For the luminosity distance to the cluster, 
we use H$_o$=50 \kms, and $\Lambda$ = 0, which gives a distance of 
2950 Mpc corresponding to a distance   modulus of
(m-M) = 42.3. Applying a  K-correction of 0.6 mag appropriate for
the cluster, the derived total luminosity of all the galaxies
is 5 $\times$ 10$^{11}$ L$_\odot$. This implies a mass-to-light 
ratio of $\sim$1200 within the assumed radius of the Einstein ring.

{\section{Discussion}}

RX J1347.5-1145 is the most luminous X-ray cluster and is one of 
the most massive of all clusters. We have used the STIS images
and the spectroscopic information of the arcs to derive a mass of 
6.3 x 10$^{14}$ M$_\odot$ within the Einstein ring
(240 Kpc). The total surface mass 
within this region as derived from the X-ray observations 
is 2.1 x 10$^{14}$ M$_\odot$ (Schindler et al. 1996) to
6.8 x 10$^{14}$ M$_\odot$ (Allen, 1997).  Thus the ratio of the
gravitational mass (which we can assume to be the total mass within the
radius) to X-ray mass is between 1 and 3.
The surface {\it gas} mass within this radius,
as derived from the  X-ray observations, is
$\sim$3.5 x 10$^{13}$ M$_\odot$.
We can assume this gas to be baryons, which imply that 
baryons contribute at least $\sim$6\% to the total mass
in this central region of the cluster.
Since the the cluster is dominated by dark matter with
mass-to-light ratio of $\sim$1200, this also implies that 
at least $\sim$6\% of the {\it dark matter} in this region
is baryonic. 
It is interesting to note that Hottari et al. (1997)
have recently derived a the mass-to-light ratio of
$\sim$ 3000 for a luminous X-ray cluster, where they find
that the bulk of the X-ray emitting gas  is also rich in metals. 

From a weak-lensing model, Fischer and Tyson (1997) 
have derived a mass of  1.0 x 10$^{15}$ M$_\odot$ within a
spherical radius of 1 Mpc
(400 arcsec). As derived from the X-ray observations,
the total mass  within this
radius is 5.8 x 10$^{14}$ M$_\odot$, and the total surface
mass within this radius is  1.0 x 10$^{15}$ M$_\odot$
(Schindler at al, 1996).
This implies that the discrepancy between the gravitational
and the X-ray mass for the whole cluster is also not large.
The mass-to-light ratio, as derived from
the ratio of the gravitational mass
to the combined luminosity from all the galaxies 
within the field, is  200 for  the  whole cluster (Fisher and Tyson, 1997). 
This is similar to what is found in other large clusters such as Coma 
(Hughes, 1989), but lower than what is seen in the central region of 
RX J1347.5-1145.

In conclusion, the discrepancy between the X-ray mass and the gravitational 
mass is not large either in the central or outer part of the cluster.
The cluster is dominated by dark 
matter, and the baryonic component in the central region is 
at least  $\sim$6\%. 

{\acknowledgements} We  would like to thank Mike Potter for 
assistance with the spectral data reduction. We  
thank Stella Seitz and Sabine Schindler for useful comments
on the manuscript.

\clearpage

\begin{table*}
\begin{flushleft}
{\caption[]{Log of the ground-based spectroscopic observations}}
\begin{tabular}{llcccccc}
\\
\hline
\\
Position & Date of & Grating   &  Integration  &Wavelength
 &Resolution   & \\
& Observation&  &  time (sec) &
range (\AA ) &(\AA )  & \\

\hline
\\

Northern arc& June 28, 1995 & R300 & 6000 & 6200 - 9600 & 8 &\\
Northern arc& June 28, 1995 & B300 & 6000 & 3600 - 6800 & 8 &\\
Southern Arc& June 29, 1995 & R300 & 6300 & 6200 - 9600 & 8 &\\
Southern Arc& June 29-30, 1995 & B300 & 7200 & 3600 - 6800 & 8 &\\
2 central galaxies& June 28, 1995 & R300 & 3600 & 6200 - 9600 & 8 &\\

\\
\hline
\end{tabular}
\end{flushleft}
 
\end{table*}

\begin{table*}
\begin{flushleft}
{\caption[]{Details of the arcs as seen in the STIS images.}}
\begin{tabular}{cccrcccccc}
\\
\hline
\\
Arc & Length  (L)$^\star$ & Width (W)$^\star$ & L/W & Distance from the & Arc Location &\\
& (arcsec)&  (arcsec)& &  central galaxy (arcsec)& RA (2000) & Dec (2000) &\\
 
\hline
\\

 1   & 5.85 & 0.76 &  7.7  & 34.2 & 13:47:32.03 & --11:44:42.1 \\
 2   & 2.09 & 0.33 &  6.4  & 31.6 & 13:47:31.13 & --11:44:38.7 \\
 3   & 2.87 & 0.33 &  8.8  & 40.2 & 13:47:27.89 & --11:45:08.5 \\
 4   & 7.81 & 0.50 & 15.6  & 36.3 & 13:47:29.26 & --11:45:39.3 \\
 5   & 1.87 & 0.20 &  9.2  & 46.0 & 13:47:31.81 & --11:45:51.7 \\

\\
\hline
\end{tabular}
\smallskip
$^\star$The STIS point spread function is 0.085 arcsec FWHM
for the  CCD and the clear aperture. 
\end{flushleft}
 
\end{table*}

\begin{deluxetable}{lcccccccc}
\small
\tablenum{3}
\tablewidth{0pt}
\tablecaption{V Magnitudes of the galaxies and the arcs$^\star$}
\tablehead{
\colhead{Filter}     & 
\colhead{Eastern Galaxy}     & 
\colhead{Central Galaxy}     &
\colhead{Arc 1}            & 
\colhead{Arc 2}            &
\colhead{Arc 3}            &
\colhead{Arc 4}            &
\colhead{Arc 5}            &                
\colhead{(V-I)}
\\
\colhead{ }            & 
\colhead{ }            &
\colhead{ }            & 
\colhead{ }            &
\colhead{ }            &
\colhead{ }            &
\colhead{ }            &
\colhead{ }} 
\startdata
Clear &19.6 & 19.2 & 22.2 & 22.2 & 23.3 & 21.5 & 23.2 & 1.0 assumed \nl
Long-pass &\nodata & 19.2$^{\dag}$  & \nodata & 22.3$^{\ddag}$  & \nodata & \nodata & \nodata & 0.93$^{\dag}$, 1.4$^{\ddag}$ \nl
\tablenotetext{\dag}{The measured color term for the brightest galaxy in 
the field.  No color information is available for the eastern CD galaxy.} 
\tablenotetext{\ddag}{The measured color term for arc 2.  This is the only 
arc for which color information can be measured.} 
\tablenotetext{\star}{The uncertainties in the magnitudes are:
0.2 for the two galaxies, and 0.4 for the arcs.}

\enddata
\end{deluxetable}
\clearpage

\figcaption[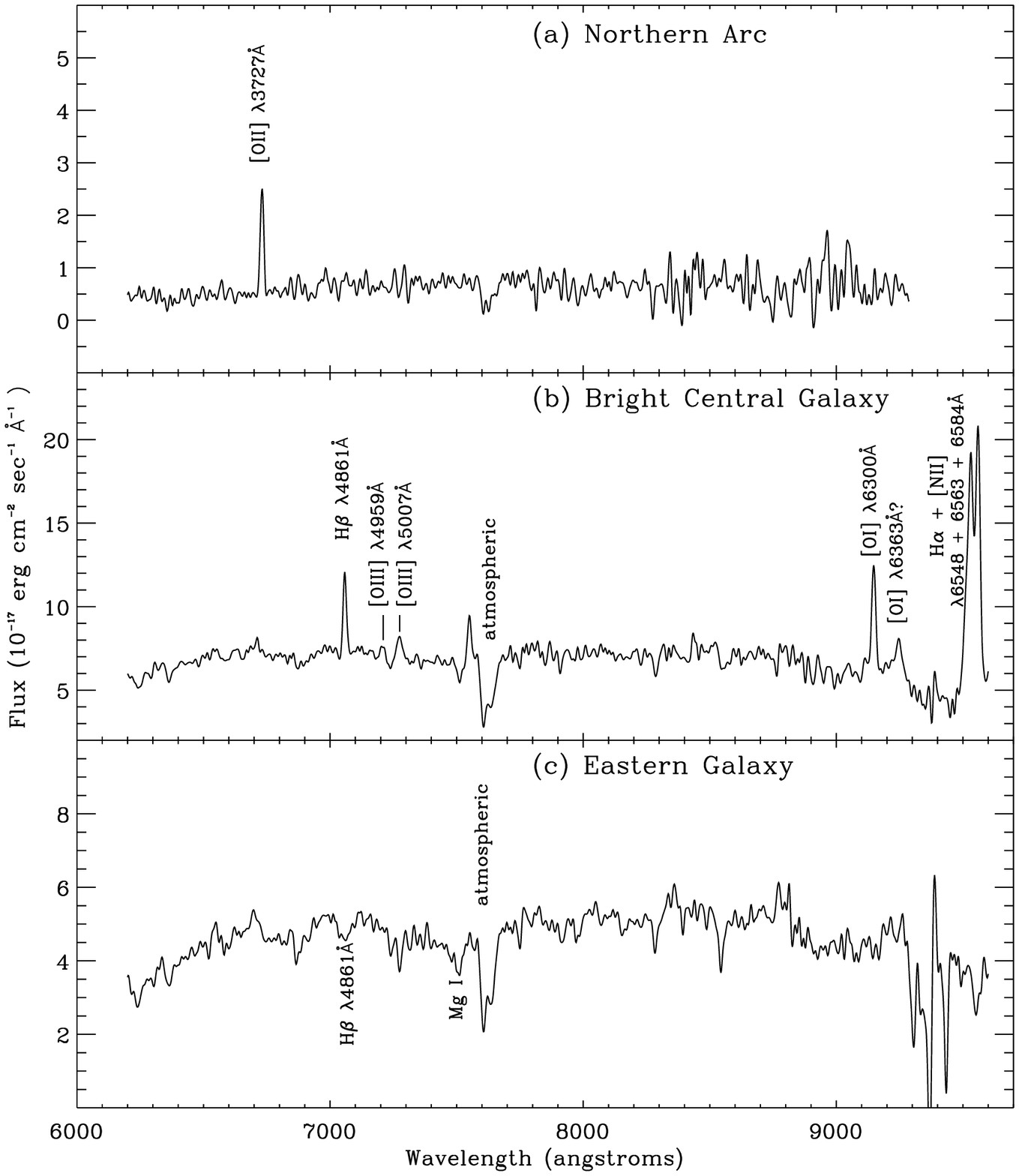]{ The spectra of the northern arc and the 
two galaxies. The region beyond 9000 \AA \  is heavily affected by 
atmospheric lines. The weak features near 9000 \AA \ in case of the
arc and the fainter galaxy are most likely due to the noise in sky
subtraction. The arc shows a strong emission line at 6730 \AA,
the central galaxy shows many emission lines, and the eastern
galaxy shows mostly absorption features.
\label{fig3}} 

\figcaption[fig2.ps]{ (Plate 1) The STIS mosaic image of the cluster 
RX J1347.5-1145 taken with the CCD and the clear filter.
The central galaxy was placed  in the overlapping region of
all the images to increase the S/N. 
The two large arcs (1 and 2) and other possible arcs
are marked in the figure.
\label{fig1}} 

\figcaption[fig3.ps]{ (Plate 2) An enlarged view (6.2$\arcsec \times 
6.2\arcsec$) of the central galaxy, which shows
a clear jet-like structure on one side, and
a similar but much fainter structure on the opposite side.
These structures, which may be jets or tidal tails, are about 
0.6$\arcsec$ long, which corresponds to a length of $\sim$ 3 kpc.
\label{fig2}}

\clearpage

\enddocument